\pdfoutput=1
\documentclass[iop]{emulateapj}
\usepackage{hyperref}

\usepackage[usenames,dvipsnames]{color}
\usepackage{rotating}
\usepackage{epstopdf}
\usepackage{color}
\usepackage{multirow}
\usepackage[para,online,flushleft]{threeparttable}
\usepackage{blindtext}

\newcommand{\halpha}{H$\alpha$}

\newcommand{\lsim}{\raise0.3ex\hbox{$<$}\kern-0.75em{\lower0.65ex\hbox{$\sim$}}}

\newcommand{\msun}{M$_{\odot}$}

\newcommand{\kms}{km\ s$^{-1}$}

\newcommand{\HI}{\mbox{\normalsize H\thinspace\footnotesize I}}
\newcommand{\HII}{\mbox{\normalsize H\thinspace\footnotesize II}}

\newcommand{\E}[1]{\ensuremath{\times 10^{#1}}}

\usepackage{fancyheadings}
\usepackage{amsmath}
\usepackage{amssymb}
\usepackage{subfigure}
\usepackage{graphics}
\usepackage{comment}
\usepackage{natbib}
\bibpunct{(}{)}{;}{a}{}{,}
\usepackage{appendix}

\defcitealias{Besla:2012jc}{B12}

\begin{document}
\title{Models of Tidally Induced Gas Filaments in the Magellanic Stream}

\shorttitle{Tidally induced filaments in the Magellanic Stream}
\shortauthors{PARDY ET AL.}

\author{Stephen A. Pardy\altaffilmark{1},
Elena D'Onghia\altaffilmark{1,}\altaffilmark{3}, 
Andrew J. Fox\altaffilmark{2}}

\altaffiltext{1}{Department of Astronomy, University of Wisconsin, 475 North Charter
  Street, Madison, WI 53706, USA (spardy@astro.wisc.edu)}

\altaffiltext{2}{Space Telescope Science Institute, 3700 San Martin Drive, Baltimore MD 21218}

\altaffiltext{3}{Center for Computational Astrophysics, Flatiron Institute, 162 Fifth Avenue, New York, NY 10010, USA}

\begin{abstract}

The Magellanic Stream and Leading Arm of HI that stretches from the Large and Small Magellanic Clouds (LMC and SMC) and over 200$^{\circ}$ of the Southern sky is thought to be formed from multiple encounters between the LMC and SMC. In this scenario, most of the gas in the Stream and Leading Arm is stripped from the SMC, yet recent observations have shown a bifurcation of the Trailing Arm that reveals LMC origins for some of the gas. Absorption measurements in the Stream also reveal an order of magnitude more gas than in current tidal models. We present hydrodynamical simulations of the multiple encounters between the LMC and SMC at their first pass around the Milky Way, assuming that the Clouds were more extended and gas rich in the past. Our models create filamentary structures of gas in the Trailing Stream from both the LMC and SMC. While the SMC trailing filament matches the observed Stream location, the LMC filament is offset. In addition, the total observed mass of the Stream in these models is underestimated of a factor of four when the ionized component is accounted for. Our results suggest that there should also be gas stripped from both the LMC and SMC in the Leading Arm, mirroring the bifurcation in the Trailing Stream. This prediction is consistent with recent measurements of spatial variation in chemical abundances in the Leading Arm, which show that gas from multiple sources is present, although the nature is still uncertain.

\end{abstract}

\section{Introduction}
\label{sec:intro}

The Magellanic System consists of the $\sim$200$^{\circ}$\HI\ Stream (\citealt{Mathewson:1974ds, Bruns:2005, Nidever:2008cz, Nidever:2010bz}), the Magellanic Bridge linking the Magellanic Clouds (``the Clouds'' or MCs) \citep{Bruns:2005}, and the Leading Arm (LA) \citep{Putman:1998ko} (for a recent review on the system see \citealt{DOnghia:2015tc}). The prime source of these features are thought to be the Milky Way's two largest companions - the Large and Small Magellanic Clouds (LMC and SMC respectively). 

The original models of the Magellanic Stream (``the Stream'' or MS) formation considered a long-term orbit with multiple tidal stripping events between the Milky Way (MW) and the LMC \citep{Toomre:1972vp, Tremaine:1975cu, Fujimoto:1976un, Fujimoto:1977ud, Lin:1995ge}. Early alternative models also focused on ram pressure stripping as the Clouds pass repeatedly through the gaseous halo of the MW \citep{Moore:1994wx}. Detailed proper motion measurements \citep{Kallivayalil:2006uu} now show that the Clouds are recent additions to the Local Group (e.g., \citealt{Besla:2007kg, Kallivayalil:2013vq}). With only one or two passes around the MW, the classic tidal interaction or ram pressure stripping are less plausible processes for gas removal. Instead, the recent infall scenario has lead to dwarf-dwarf galaxy interaction models where the MW plays only a limited role \citep{Besla:2010kg, Diaz:2011bb}. The current best model by \citet{Besla:2012jc}  (B12, hereafter), which has the LMC and SMC interacting before a recent infall to the MW's virial radius, can reproduce many of the observed Stream characteristics. This model predicts that the Stream was stripped primarily from the SMC; a common feature among studies examining the chemical enrichment \citep{Fox:2013kc} and dynamics \citep{Yozin:2014ti} of the Stream. Nevertheless, the MS is bifurcated in both kinematic \citep{Connors:2006hc, Nidever:2008cz} and chemical space \citep{Gibson:2000cn, Richter:2013hk}, with one of the filaments connecting back to the LMC, a feature not produced in the B12 models. This filament shows that gas from the LMC contributed to the production of the Stream.

Beyond \HI, recent observations have found evidence for many small stellar streams in the vicinity of the Clouds \citep{Besla:2016wh}, some of which overlap the MS \citep{Belokurov:2015um}, and some, especially near the Bridge, which appear to be stripped from the LMC \citep{Belokurov:2016tj, Deason:2016ar}. Nevertheless, observational studies have failed to find a stellar counterpart to the Stream which would be predicted by tidal models \citep{RecillasCruz:1982bu, Brueck:1983wl}. The Stream also contains copious ionized hydrogen, first detected in \halpha\ by \citet{Weiner:1996fc, Putman:2003hf}. More recently, the Wisconsin \halpha\ Mapper (WHAM) map of the Bridge found that up to 50\% of the gas is ionized \citep{Barger:2013cv}. In the Trailing Stream, these ionization fractions can be above 90\% \citep{Fox:2014ih, Barger:2017wr}. Adding the total neutral and ionized gas gives a Stream mass of 2.0\E{9}\msun; larger than the total present-day gas mass of the LMC and SMC combined (8.43\E{8} \citealt{Bruns:2005}) \citep{Fox:2014ih}, as well as the \HI\ Stream (as predicted by \citealt{BlandHawthorn:2007kc}). If the mass budget of the Stream is confirmed it suggests that the Clouds were much richer in gas in the past and the subsequent stripping  efficiency was extremely high, a characteristics not current accounted for the dwarf-dwarf galaxy interaction models. Indeed, the current dwarf-dwarf galaxy models tend to \emph{over-predict} the mass in the Clouds, while \emph{under-predicting} the mass in the Stream (\citetalias{Besla:2012jc}, \citealt{Guglielmo:2014bg}).

This work aims to reproduce the observed filaments and mass content in the MS by tidal interaction between the SMC and LMC at the first infall. This requires that the Clouds were more gas rich in the past and that the stripping efficiency of the gas from the LMC stripping was much higher. To facilitate these changes, we run simulations of encounters between the LMC and SMC with varying mass ratios and gaseous disk scale lengths. In this paper, we focus on a simulation in which, keeping all other factors equal, we drastically increased the size of the LMC gaseous disk and total gas mass of both dwarfs.

In the following section we will lay out our simulation tools and initial conditions. We will then dive into our simulation results in \autoref{sec:results} by showing the dwarf-dwarf galaxy interaction before the Clouds fall into the MW potential in \autoref{subsec:premw} and after they have entered the halo in \autoref{subsec:mw}. \autoref{sec:discussion} summarizes the main results of this work and discusses the implications.

\section{Numerical Methods}
\label{sec:method}

We ran all of our simulations using the parallel TreePM-Smoothed Particle Hydrodynamics (SPH) code GADGET3. In this code, stars and dark matter are treated as collisionless fluids, while gas particles include hydrodynamical forces. 

The most recent description of the code is available in \citet{Springel:2005cz}. The GADGET3 code incorporates a sub-resolution multiphase model of the ISM including radiative cooling \citep{Springel:2003eg} and a fully conservative approach to integrating the equations of motion \citep{Springel:2002ef}. All of our simulations include full gas physics, including radiative cooling and density-based star formation. This work, however, is not designed to test the hydrodynamical affects on the properties of the MS.

One important caveat of our simulation setup is that the MW is modeled with a static NFW dark matter halo. This affects three aspects of the simulations: the orbit of the LMC is known to vary based on whether the MW is fixed to the center of the simulation grid \citep{2015ApJ...802..128G}; it does not include a disk for the MW, excluding any possible interaction of the disk with the LA; and it lacks a hot gas halo, excluding any possible ram pressure stripping. These choices, along with many others throughout this work, were made to allow for easy comparison with \citetalias{Besla:2012jc}.

\subsection{Initial Conditions}
\label{subsec:ICs}

We generated equilibrium models of the LMC and SMC in isolation using the methods outlined in \citet{Springel:2005co}. The gas and stellar disk component of each galaxy is a thin exponential surface density profile of scale length $r_{d}$, scale height of $z_{0}$, and total disk mass of M$_{disk}$ such that the volume density is given by:
\begin{equation}
\rho_{\star}(r,z)= \frac{M_{disk}}{4\pi z_{0}r_{d}^2} \rm{sech}^2 \Big(\frac{z}{z_0}\Big) exp\Big(-\frac{r}{r_{d}}\Big)
\end{equation}

The scale height of the stellar and gaseous disk is initially adopted as 20\% of the disk scale length. The energy and pressure of the ISM is prescribed by the chosen effective equation of state \citep{Springel:2005co}.
Our choice of disk parameters for the LMC and SMC are taken from \citet{2002AJ....124.2639V} and \citet{Stanimirovi:2004ew} respectively and motivated by the work by \citetalias{Besla:2012jc} (their table 1). We show all of our model parameters in \autoref{tab:ic_params}. Columns 2-4 of \autoref{tab:ic_params} contain the scale lengths of the dark matter, stellar, and gaseous components of each galaxy, columns 5-7 contain the masses of each component, and columns 8-10 contain the number of particles. We used a softening length of $290$ pc for the dark matter, and $100$ pc for the stars and gas particles.

We use a Hernquist \citep{Hernquist:1990hf} halo for the dark matter distribution of each galaxy, given by:
\begin{equation}
\rho(r) = \frac{M_{DM}}{2\pi}\frac{a}{r(r+a)^{3}},
\end{equation}  
where $M_{DM}$ is the galaxy halo mass and $a$ is the scale length of the halo. The inner parts of the Hernquist halo also approximates those of the NFW halo \citep{1997ApJ...490..493N}. Conversion between the two halos can be accomplished with a fitting formula given as eqn. 2 of \citet{Springel:2005co}.

The choice of halo parameters for the LMC and SMC before the present day is not well constrained by observations. Although stellar abundance matching masses of the two Clouds gives mass ratios as low as 2:1 for the Clouds \citep{2017MNRAS.472.1060D, Dooley:2016vt}, the dwarf-dwarf galaxy interaction model requires that the LMC and SMC remain a long-term binary pair, a feat more difficult with a lower mass ratio. We fix the halo concentrations ($c = r_{200}/r_s$, where $r_s$ is the scale length of an equivalent NFW halo) as 15 for the SMC and 9 for the LMC, as done in \citetalias{Besla:2012jc}. We then let the halo scale lengths vary as we increase or decrease the total mass of the SMC and LMC.

Our fiducial case has an SMC and LMC with large gas fractions and with gas disk scale lengths that are 4 times larger than their stellar disk scale lengths, and with a 9:1 mass ratio (chosen to be the same as in \citetalias{Besla:2012jc}).  We refer to this simulation as the 9:1 model throughout the paper. Dwarf galaxies are known to have very extended disks \citep{Swaters:2002cg, Kreckel:2011ff}, and our 9:1 simulation is a purposefully extreme example to show what is required to strip material efficiently from the LMC. 

We have also conducted runs using 3:1, 5:1, and 7:1 mass ratios which give broadly consistent results. We note the similarities and differences in \autoref{subsec:massratio}. 

We caution the reader that a full parameter space search is not possible for these models given the computational costs. Instead, for each simulation setup, we run on order $\sim$10 simulations -- varying orbital parameters and selecting those which best match the observed Stream location. 

\begin{table*}[htbp]
   \centering
\caption{STRUCTURAL PARAMETERS OF THE PRIMARY AND COMPANION GALAXY}
      \begin{threeparttable}
   \begin{tabular}{@{} lccccccccr @{}}
      \hline
      Galaxy & a (kpc)\tnote{a} & R$_{\mathrm{stars}}$ (kpc)\tnote{b} & R$_{\mathrm{gas}}$ (kpc)\tnote{c} & M$_{\mathrm{Halo}}$ ($\times10^{10}$M$_{\odot}$)\tnote{d} & M$_{\mathrm{Stars}}$ ($\times10^{10}$M$_{\odot}$)\tnote{e} & M$_{\mathrm{Gas}}$ ($\times10^{10}$M$_{\odot}$)\tnote{f} & N$_{\mathrm{Halo}}$\tnote{g} & N$_{\mathrm{Gas}}$\tnote{h} & N$_{\mathrm{Disk}}$\tnote{i} \\
      \hline
LMC extended & 21.7 & 1.8 & 5.4 & 17.64 & 0.25 & 0.1 & 1000000 & 300000 & 600000  \\ 
LMC More Gas & 21.7 & 2.4 & 9.5 & 17.53 & 0.25 & 0.2 & 1000000 & 500000 & 600000  \\ 
SMC 9:1 & 7.3 & 1.2 & 4.7 & 1.92 & 0.03 & 0.2 & 109000 & 359000 & 62000  \\ 
SMC 7:1 & 7.7 & 1.3 & 3.8 & 2.40 & 0.03 & 0.1 & 143000 & 278000 & 81000  \\ 
\hline
\end{tabular}
\begin{tablenotes}

\item[a] Dark matter halo scale length.
\item[b] Stellar disk scale length
\item[c] Gaseous disk scale length.
\item[d] Dark matter halo mass.
\item[e] Stellar disk mass.
\item[f] Gaseous disk mass.
\item[g] Number of dark matter particles.
\item[h] Number of gas particles.
\item[i] Number of stellar particles.
\end{tablenotes}
\end{threeparttable}
\label{tab:ic_params}
\end{table*}

\subsection{Orbital Configurations}
\label{subsec:orbits}

Our simulation follows the setup of \citetalias{Besla:2012jc} and is motivated by the proper motion measurements and past orbits of the LMC and SMC from \citet{Kallivayalil:2013vq} where, due to them being on their first infall, the influence of the MW potential has been minimal and only very recent. We design simulations to remove gas from both the LMC and SMC by dwarf-dwarf galaxy interaction before the Clouds fall into the MW. We set up our simulations such that the Clouds first interaction with each other over a period of 4-6 Gyrs in isolation. We then stop the simulation and place them under the influence of a fixed potential for the MW. The use of a first passage model and a fixed MW halo is known to be inaccurate for certain choices of LMC and MW mass \citep{2015ApJ...802..128G}. Nevertheless, we use this methodology as a way to directly compare our findings with the results of \citetalias{Besla:2012jc}. 

\begin{figure}[htbp] 
   \centering
   \includegraphics[width=3in]{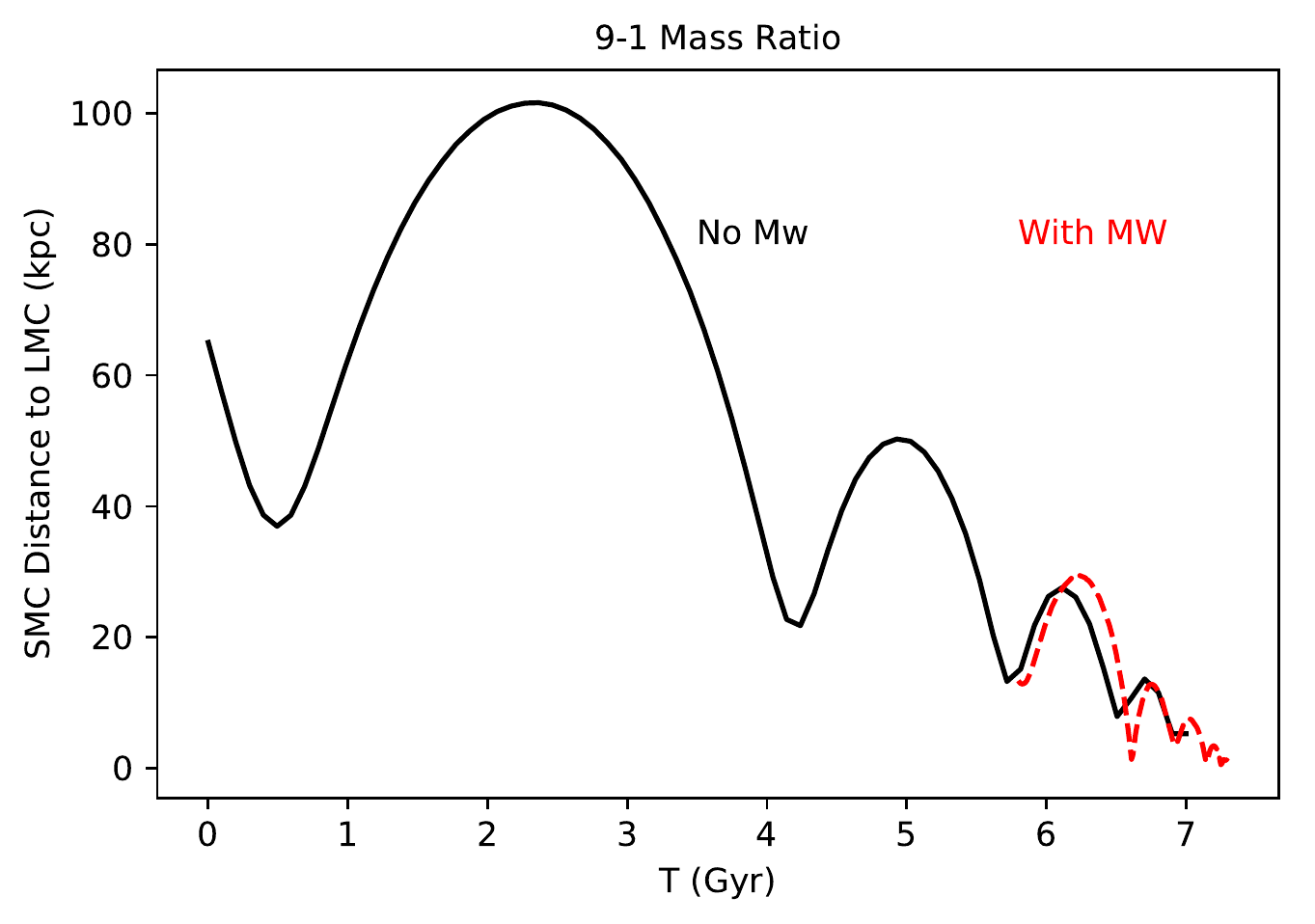} 
   \caption{Orbit of the SMC around the LMC in the 9:1 mass ratio model. The orbit before entering the MW potential is shown as a black solid line, and the orbit after entering the MW potential is shown with a red dashed line.}
   \label{fig:orbit_premw}
\end{figure}

Our dwarf-dwarf galaxy interaction was set such that the SMC had an initial distance to the LMC of 65 kpc on an eccentric orbit (e=0.65), and with an initial pericenter distance of 25 kpc. The exact starting positions and velocities are given in \autoref{tab:orbits}. The orbit is in-plane, but the disk-planes of each galaxy are tilted with respect to the orbit plane. The LMC is first rotated by 90$^{\circ}$ around the x-axis and then by 45$^{\circ}$ around the z-axis. The SMC is rotated by 90$^{\circ}$ around the y-axis. The orbit of the SMC around the LMC for the 9:1 mass ratio model can be seen in \autoref{fig:orbit_premw}. The orbit of the SMC around the LMC with no influence of the MW potential is shown as a solid black line, while the orbit of the SMC around the LMC under the influence of the MW potential is shown as a dashed red line. In our model, the SMC completes three encounters with the LMC before the simulation is halted and the Clouds are placed into the MW potential (see \autoref{subsubsec:mworbits}). The other mass ratio models are designed to have a very similar orbit.

\begin{table}[htbp]
   \centering 
   \caption{ORBITAL PARAMETERS OF DWARF-DWARF INTERACTION}
      \begin{threeparttable}
   \begin{tabular}{@{} lcccccr @{}} 
      \hline
      Model & x & y & z & V$_x$ & V$_y$ & V$_z$  \\
                 & \multicolumn{3}{c}{(kpc)} & \multicolumn{3}{c}{(\kms)} \\
      \hline
9:1 Mass Ratio & 36.6 & 54 & 0 & $-$120.8 & $-$13 & 0 \\
7:1 Mass Ratio & 36.5 & 54 & 0 & $-$121.6 & $-$13 & 0 \\
5:1 Mass Ratio & 36.5 & 54 & 0 & $-$124.8 & $-$13 & 0 \\
3:1 Mass Ratio & 22.1 & 72 & 0 & $-$126.7 & $-$54 & 0 \\
      \hline
   \end{tabular}
           \begin{tablenotes}
        	  Note: The origin of the coordinate system is at the center-of-mass of the LMC.
        \end{tablenotes}
     \end{threeparttable}
   \label{tab:orbits}
\end{table}

\subsubsection{Milky Way Infall}
\label{subsubsec:mworbits}

After three passes of the SMC around the LMC, the Clouds are placed at 220 kpc from the MW's center (roughly the virial radius) with a relative velocity of 163 \kms\ for the LMC and $\sim$300 \kms\ for the SMC. We attempted to pick initial conditions that best matched the present-day positions of the LMC, SMC, and Stream. The final positions and velocities of the Clouds are highly sensitive to their initial conditions, and the simulations are sufficiently expensive to preclude a full parameter space search.

The 9:1 mass ratio simulation completes the three passes in 5.4 Gyrs and then falls to the Clouds' current positions in 1 Gyr. The other mass ratio interactions are modified to have similar interaction and infall times.

\begin{table}[htbp]
	\label{tab:orbit_mw}
   \centering 
   \caption{ORBITAL PARAMETERS DURING MILKY WAY INFALL}
      \begin{threeparttable}
   \begin{tabular}{@{} lcccccr @{}} 
      \hline
      Model & x & y & z & V$_x$ & V$_y$ & V$_z$  \\
                 & \multicolumn{3}{c}{(kpc)} & \multicolumn{3}{c}{(\kms)} \\
	9:1 Mass Ratio & & & & & \\
	LMC & $-$22.0 & 217 & 32 & 22.1 & $-$120 & $-$109 \\
	SMC & $-$29.9 & 217 & 22 & 42.9 & $-$284 & $-$80 \\
	\hline
	7:1 Mass Ratio & & & & & \\
	LMC & $-$32.4 & 200 & 86 & 18.6 & $-$88 & $-$137 \\
	SMC & $-$30.9 & 195 & 75 & 119.2 & $-$216 & $-$123 \\
	\hline
	5:1 Mass Ratio & & & & & \\
	LMC & $-$21.7 & 198 & 95 & 12.4 & $-$83 & $-$140 \\
	SMC & $-$26.1 & 199 & 89 & 40.7 & $-$260 & $-$163 \\
	\hline
	3:1 Mass Ratio & & & & & \\
	LMC & 47.3 & 197 & $-$86 & $-$16.4 & $-$160 & $-$28 \\
	SMC & 51.5 & 206 & $-$101 & $-$48.2 & $-$282 & 22 \\
      \hline
   \end{tabular}
           \begin{tablenotes}
        	  Note: The origin of the coordinate system is the center of the MW.
        \end{tablenotes}
     \end{threeparttable}
\end{table}

\subsection{The Magellanic Stream Mass}
\label{subsec:tidalstripping}

We measured the gas stripped from the LMC and SMC by computing when gas particles were no longer bound to the main body of their host galaxy. For each gas particle in our simulation, we calculate the gravitational potential and its kinetic energy. Any particle that has a larger kinetic than potential energy is considered unbound. We then make mock \HI\ images of the Stream by projecting the gas particle locations into Magellanic Coordinates and binning in 35 arcminute bins. We then compute the column density along each bin and blank any bins where the column densities are below 5.2\E{17} cm$^{-2}$. This produces a roughly comparable map to \citet{Nidever:2010bz}. From this map, we then sum the total mass, assuming a fixed distance of 50 kpc, as commonly done in the literature. 

\section{Results}
\label{sec:results}

We present the formation of the Stream by gas stripped from both the LMC and SMC by mutual gravitational encounters at the first infall. We first discuss the orbits while the Clouds are in isolation (\autoref{subsec:premw}), and then discuss the influence of the MW potential (\autoref{subsec:mw}). Our primary results are presented as gas density maps of the Magellanic Clouds at various times during their orbit. 

\subsection{Dwarf-Dwarf Galaxy Interactions}
\label{subsec:premw}

In our models the Stream is produced by gas tidally stripped from both the LMC and the SMC during their repeated close encounters before entering the MW's DM halo. The majority of gas is stripped during the last 2 Gyr. The gas from the LMC in particular is stripped mostly after the Clouds enter the MW.

\begin{figure*}[htbp] 
   \centering
   \includegraphics[width=7in]{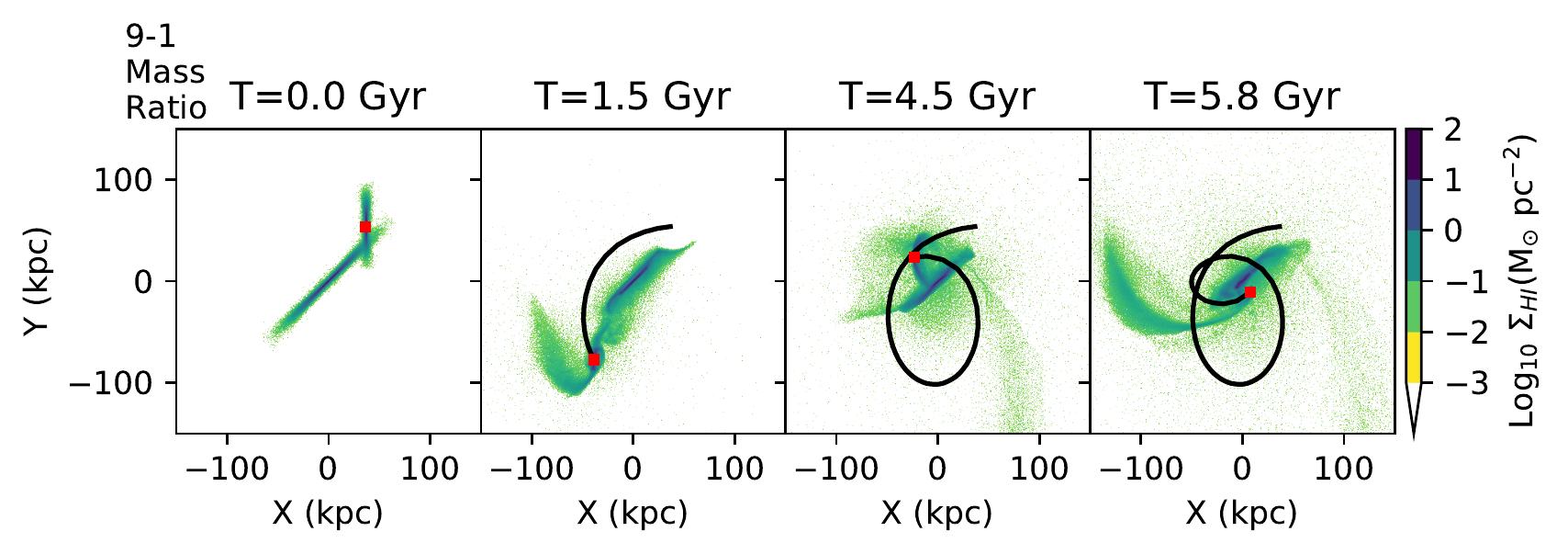} 
   \caption{Tidal interaction between SMC and LMC before infalll to the MW potential in the 9:1 LMC-SMC mass ratio model.  Panels show gas surface density in a logarithmic scale (the colorbar on the right is the same for all panels). The red square and solid black line show the center of the SMC and its orbit. The simulation is shifted to place the LMC at center of each panel. The time since the simulation started is given above each panel, and the times have been chosen to show a range of interaction configurations.}
   \label{fig:premw}
\end{figure*}

\autoref{fig:premw} shows the gas surface density of the LMC and SMC as they interact before infall to the MW potential in the 9:1 mass ratio simulation. The SMC is marked by a red square and its orbit is indicated by a solid black line. All simulations have been shifted to place the center of mass of the LMC at the center of the frame. The time of each of the four panels is shown from the start of the simulation, and are chosen to show a range of the interaction configurations. The initial pass strips material from the SMC and distorts the disk of the LMC similarly to what has been seen in \citet{Pardy:2016uu} (see the second panel at t=1.5 Gyr). As the SMC approaches for a second pass, it produces a low density tail in the lower right quadrant of the frame (third panel at t=4.4 Gyr). In its final pass, the SMC has swept out a long tail in the top and bottom left quadrants of the frame (final panel at t=5.4 Gyr) which will form the Stream after infall to the MW. This final frame is the stopping point for this simulation.

\subsection{Formation of the Magellanic Stream}
\label{subsec:mw}

\begin{figure*}[t] 
   \centering
   \includegraphics[width=7in]{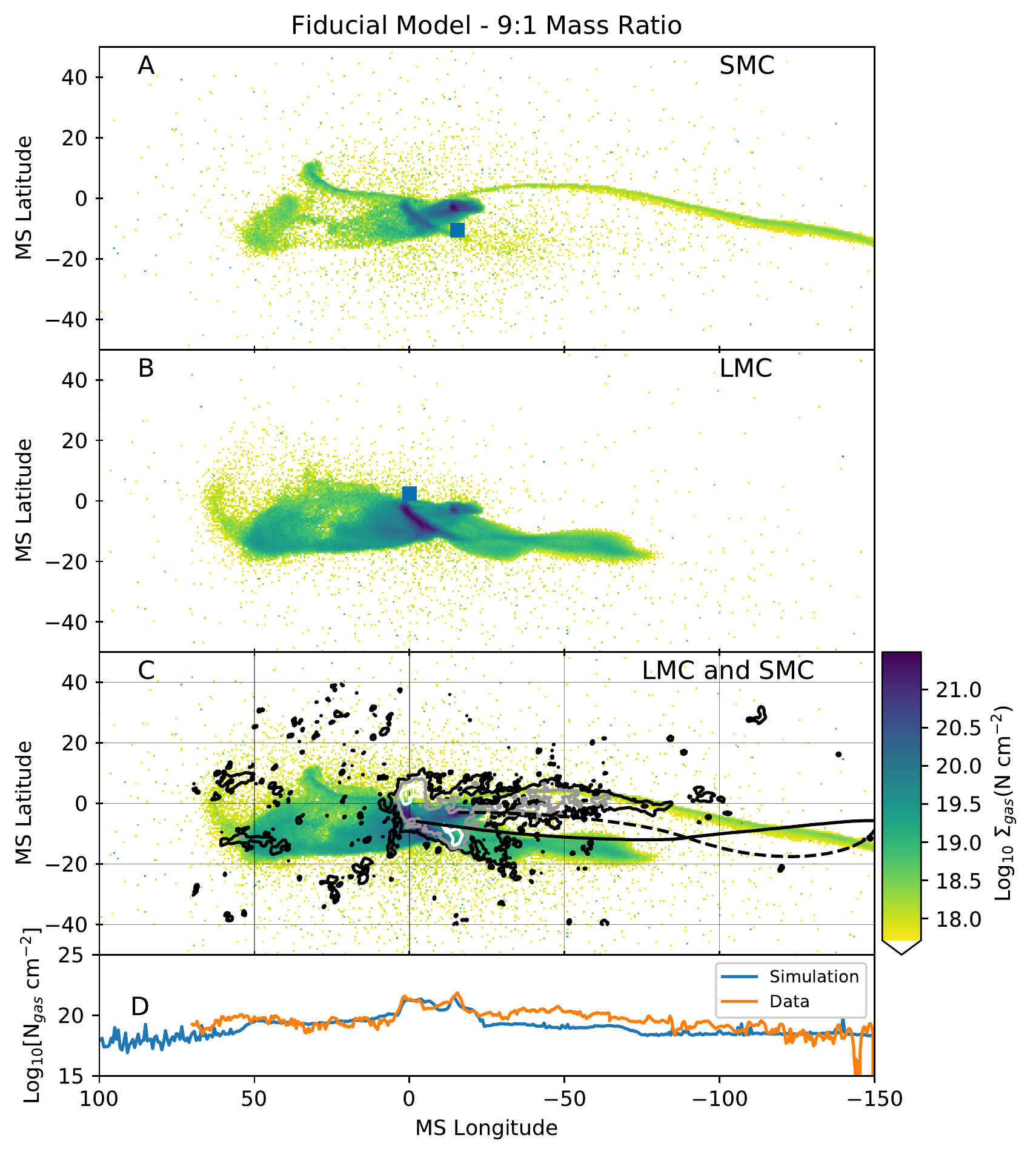} 
   \caption{The MS in gas according to our simulation with a 9:1 LMC-SMC mass ratio. The top three panels (A-C) show the column density of simulated Stream material in Magellanic Coordinates \citep{Nidever:2008cz}. Gas from the SMC and LMC are displayed individually in panels A and B respectively, along with the current observed position of the galaxies (blue squares). Panel C shows the full \HI\ Stream in our simulation. Column density is computed in 35 arcsecond bins, and densities below 5.2$\times$10$^{17}$ are blanked to make mock observations similar to those presented in \citet{Nidever:2010bz}. Over-plotted in this panel is data from \citet{Nidever:2010bz} shown in logarithmic contours of levels \{10$^{19}$ (black lines), 10$^{20}$ (gray lines), 10$^{21}$ (white lines)\} N cm$^{-2}$. Also shown on this panel are the orbit lines for the LMC and SMC since entering the MW potential. The LMC orbit is indicated by the solid black line while the SMC is indicated by the dashed black line. Panel D shows the maximum column density of \HI\ gas in both our simulated Stream and from the data.}
   \label{fig:9-1_stream}
\end{figure*}

After we place the two Clouds in the MW halo, they continue interacting for roughly 1 Gyr before we stop them at the present day. \autoref{fig:9-1_stream} shows the gas surface density of the present day Stream in the 9:1 mass ratio simulations. This figure contains four panels illustrating the details of the present day gas distribution. We show, from bottom to top: the maximum column density of \HI\ from \cite{Nidever:2010bz} compared with the maximum column density from our simulation (panel D); the gas surface density of our simulation (background colors with colorbar scale on right) compared with the \HI\ data from \citet{Nidever:2010bz} in gray contours (panel C); the gas surface density of only gas particles that originated from the LMC (scale is the same as in panel C; panel B); the gas surface density from particles that originate from the SMC (scale is the same as in panel C; panel A). By splitting the particles by origin we can trace the Stream's contributions from each galaxy. 

We note a few features in this model. First, both galaxies contribute to the Stream material, as has been found observationally from chemical abundance analyses of the two principal filaments in the Stream \citep{Fox:2010hc, Fox:2013kc, Richter:2013hk}. The greatest angular extent of material in our model is contributed by the SMC, but the LMC has a dense $\sim$50 degree stream on the trailing side and a leading feature of nearly the same size, along with a low density filament that follows the SMC filament. We match the qualitative shape of the Stream as well as the maximum column density of \citet{Nidever:2010bz} in the Leading Arm. Note that although we match the observed column densities better than in previous works, the observed density of the Trailing Stream is still a factor of 10 higher than our models, and found in the LA (see panel D of \autoref{fig:9-1_stream}). We remark that this seems to be a fundamental limitation of models of the Stream \citep{Besla:2012jc, 2015ApJ...813..110H}.

\begin{figure*}[t] 
   \centering
   \includegraphics[width=7in]{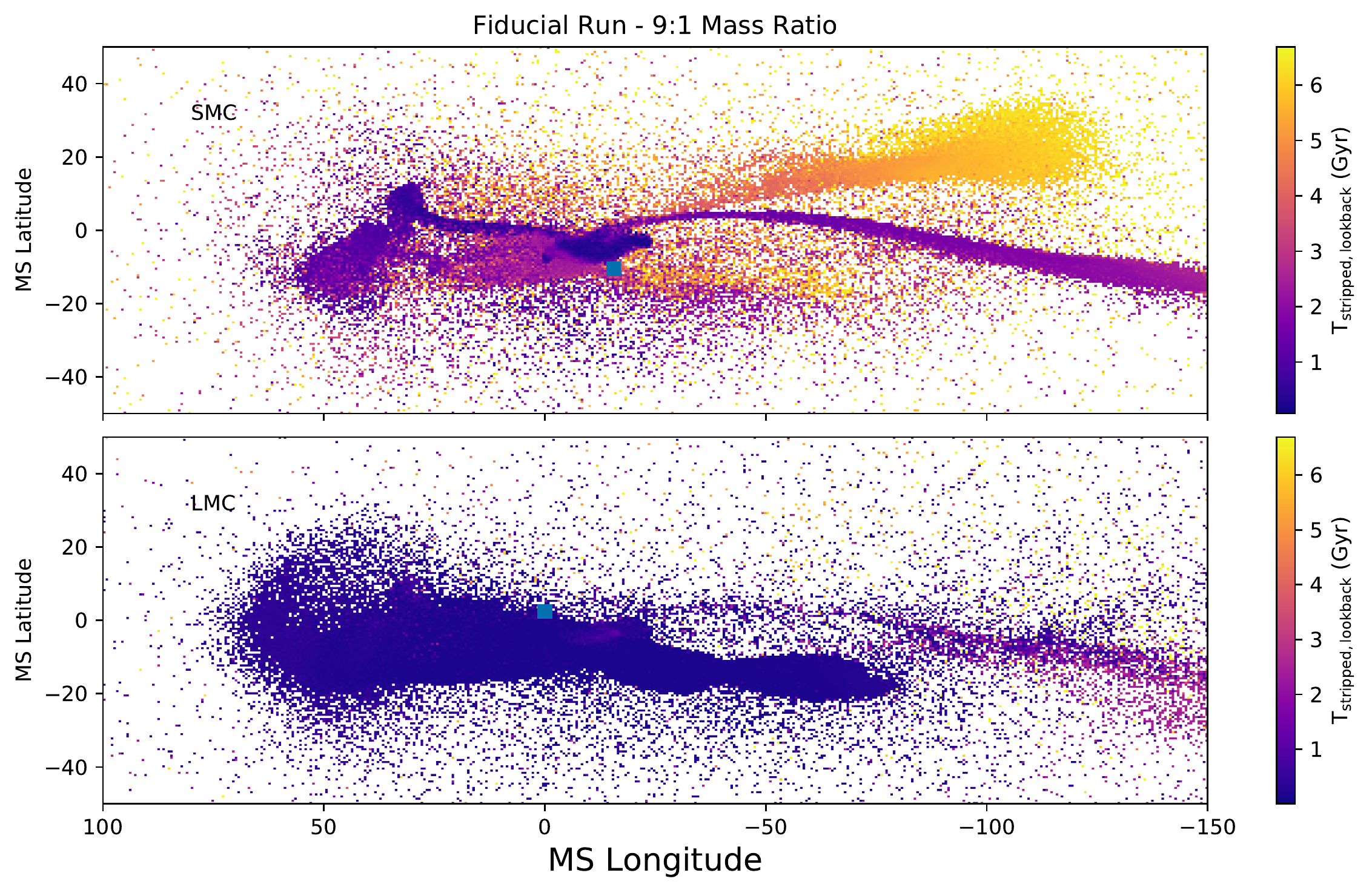} 
   \caption{The dynamical age of the Stream in the 9:1 mass ratio model color-coded by time since the gas was stripped. For each bin in \autoref{fig:9-1_stream} we compute the time since the particles were first stripped from the galaxy, with t=0 being the present day. Average ages in each bin are shown from the SMC and LMC are plotted in Magellanic Coordinates in the top and bottom panels respectively. The SMC filament is primarily composed of gas particles stripped 2-3 Gyrs ago, along with a tail of material stripped from an earlier encounter (it is uncertain if this material would survive for such a long time in the presence of a hot halo). The LMC filament, in contrast, is composed almost entirely of recently stripped material. As in other figures, solid blue squares show the present day centers of the LMC and SMC.}
   \label{fig:9-1_stripping_time}
\end{figure*}

The two galaxies do not create parallel and twisting filaments as observed in the data. Instead, the material from the LMC and SMC are in separate filaments. Much of the gaseous Stream material was stripped after the most recent collision between the Clouds when they were the closest to each other. The SMC filament is made of gas stripped in the last 2-3 Gyr from the last few interactions with the LMC and is thinner on the sky. The LMC material is almost entirely stripped within the last Gyr and produces a thick, shorter ($\sim$50$^{\circ}$) filament. We quantify the stripping time by tagging each gas particle by the time when they were first stripped from their host galaxy, then computing the average for each bin given in \autoref{fig:9-1_stream}. We show the results in \autoref{fig:9-1_stripping_time}.

Our modification of the initial mass models of the LMC and SMC give rise to present day positions that are 8.5 and 9.7 kpc, respectively, away from their observed positions. This provides a better match to the SMC's position than found in model 2 of \citetalias{Besla:2012jc}, but a worse match to the LMC's position. 

\begin{figure}[t] 
   \centering
   \includegraphics[width=3.5in]{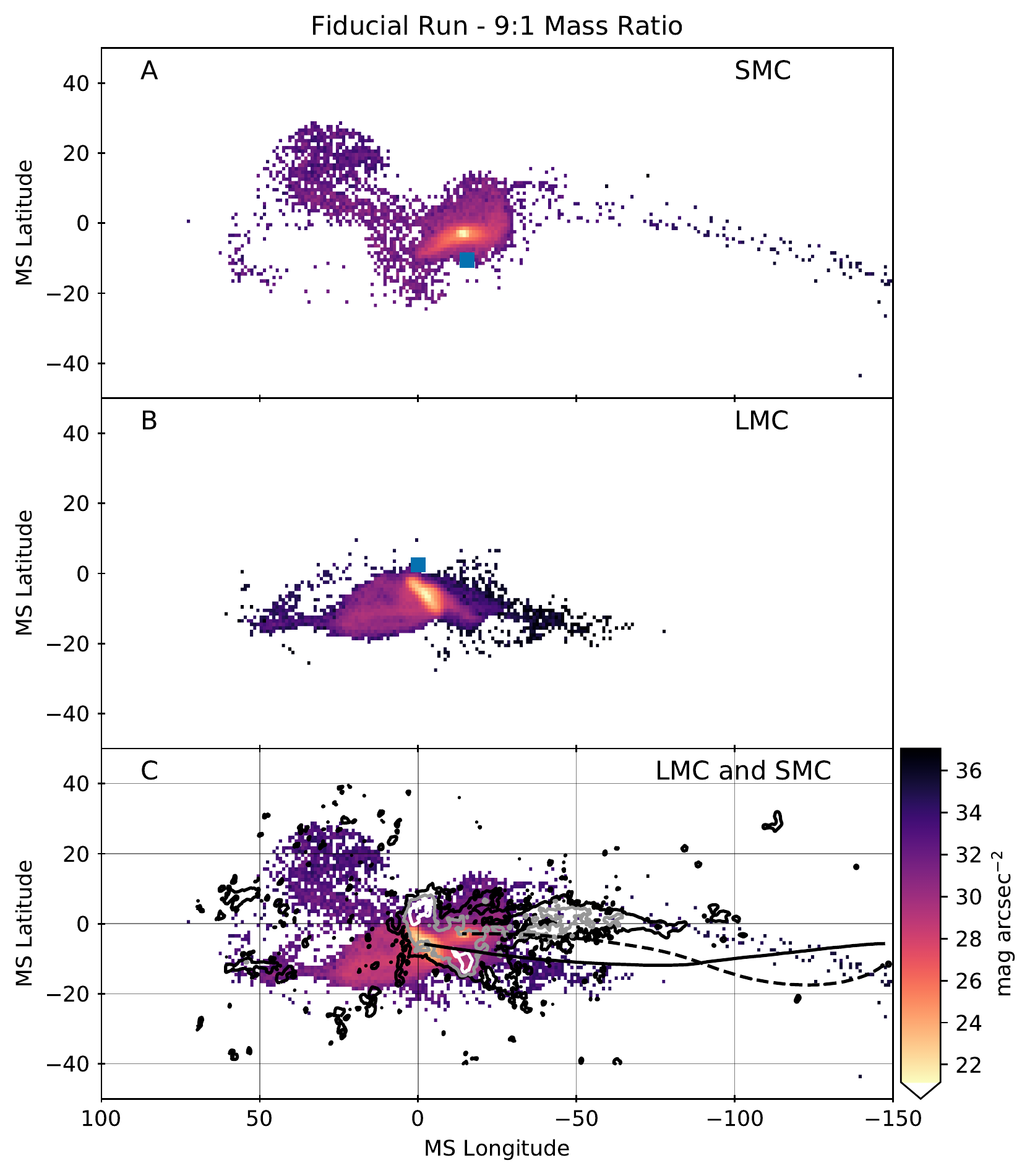} 
   \caption{The MS in the stellar component according to our simulation with a 9:1 LMC-SMC mass ratio. The panels are similar to those found in \autoref{fig:9-1_stream}, but instead of column densities we show magnitudes per square arcsecond. Blue squares in panels A and B show the present day positions of the centers of the SMC and LMC respectively. For this analysis we have assume a single age population (5 Gyr old) observed in the V band. The majority of stars are stripped from the SMC and extend in a very thin, very low density ($\sim$ 30 mag arcsec$^{-2}$) filament that would be difficult to observe.}
   \label{fig:9-1_stellar_stream}
\end{figure}

The dwarf-dwarf galaxy interaction model also strips stars from the LMC and SMC during their mutual interaction. \autoref{fig:9-1_stellar_stream} shows the surface brightness of stars in magnitudes per square arcsecond in Magellanic Stream coordinates. Again, we split the stars into population based on which galaxy they were stripped from. For this analysis we have assumed a single age population (5 Gyr old) observed in the V band. The majority of stars are stripped from the SMC and extend in a very thin, very low density ($\sim$ 30 mag arcsec$^{-2}$) filament that would be difficult to observe. There is a more significant stellar component between the two Clouds (as is seen in the observed Bridge) and close to the Clouds (see the stellar streams found in recent observations by \citealt{Belokurov:2016tj} and \citealt{Deason:2016ar}). We also find a more dense stellar component in the LA, broadly consistent with the early results from the SMASH survey \citep{Nidever:2017vf}.

To fully quantify the mass contributions to the Stream from the LMC and SMC, we measure the gas tidally stripped from each galaxy as described in \autoref{subsec:tidalstripping}. The LMC and SMC contribute 2.3, 1.6 \E{8} \msun\ to the Stream, respectively. For a full account of the initial and final gas mass in each galaxy, and the total gas stripped by tidal interactions see \autoref{tab:mass_stripping}. \emph{The total mass in our simulation is still an order-of-magnitude less than the upper limits placed on the observed Stream if the ionized gas is included.}

\begin{figure*}[htpb] 
   \centering
   \includegraphics[width=7in]{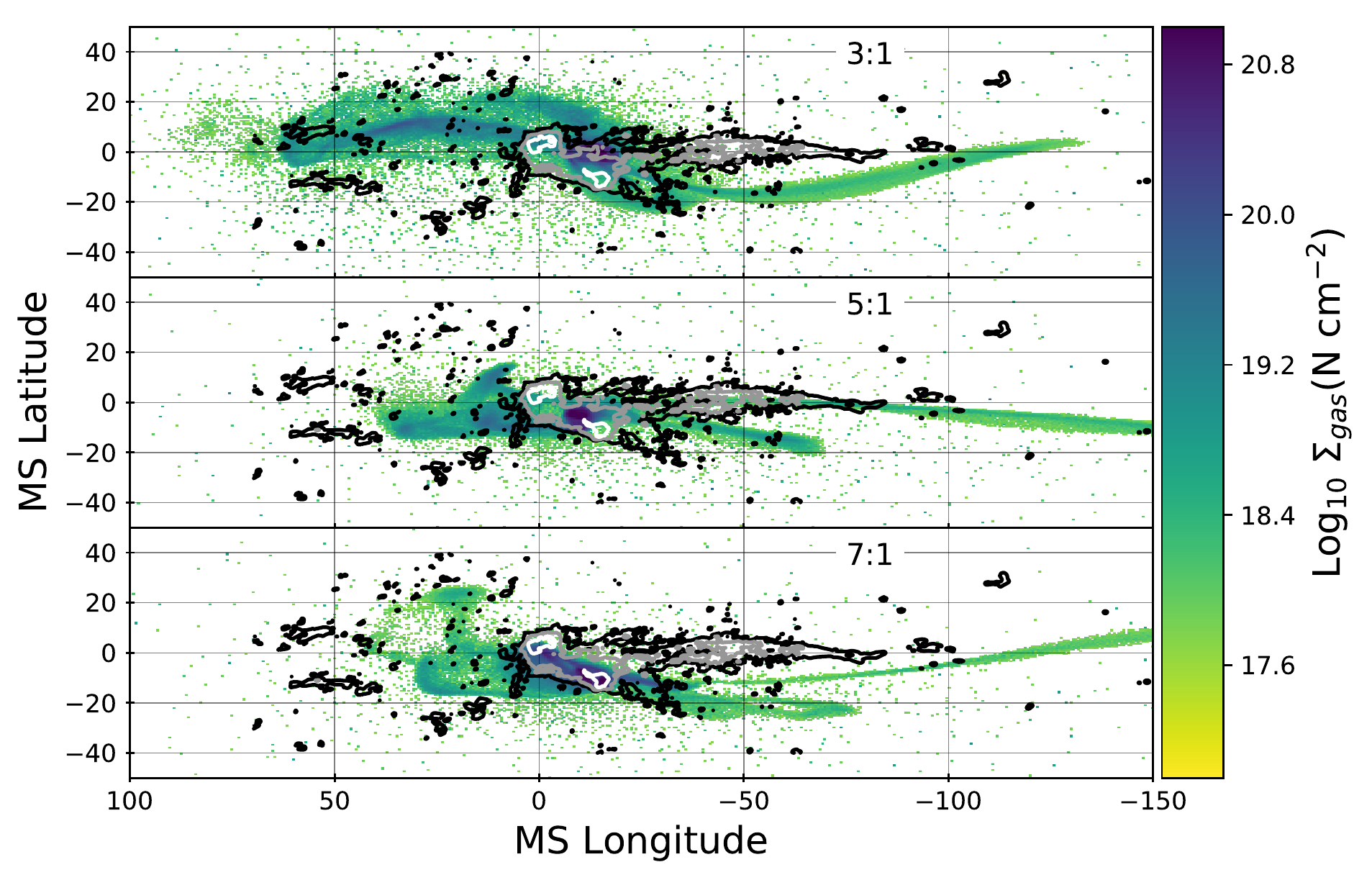} 
   \caption{Magellanic Stream coordinate representation for different mass-ratio interactions. Each panel shows the density of gas in the simulations as in panel C of \autoref{fig:9-1_stream}.}
   \label{fig:other_mass_stream}
\end{figure*}

We also examined the line of sight distances and velocities to the Stream in our two models. As is found in \citetalias{Besla:2012jc}, the Clouds are located at $\sim$ 50 kpc, as in observations, and the Stream gets further away with larger negative Magellanic Longitudes. The tip of the Stream in our fiducial model is 150-200 kpc. The LA feature bends such that its closest point to us, just more than 25 kpc, occurs about half-way along its full extent. This appears to conflict with the observations of the LA that place some of that material as close as 17 kpc \citep{McClureGriffiths:2008di, 2014ApJ...784L..37C}.

\begin{figure}[htbp] 
   \centering
   \includegraphics[width=3.5in]{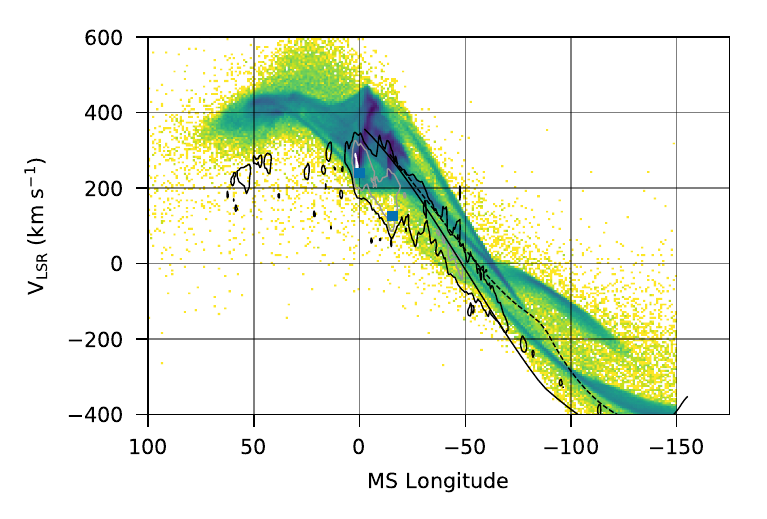} 
   \caption{The line of sight velocity (in the Local Standard of Rest frame) to the MS in the 9:1 SMC model. Colors show relative density of gas in a logarithmic scale. The \HI\ velocity gradient from \citet{Nidever:2010bz} is shown as the dark contours. The LMC orbit is shown as a solid black line while the SMC is the dashed black line. Blue squares show the position and velocities of the observed LMC and SMC (the center of LMC is currently at 0$^{\circ}$ longitude).}
   \label{fig:stream_vlos}
\end{figure}

Our simulations qualitatively reproduce the observed line of sight velocities, showing a gradient towards more negative velocities along the Stream, but are offset by $\sim$100 \kms\ with respect to the observations (see \autoref{fig:stream_vlos}). This figure also show the \HI\ velocity from \citet{Nidever:2010bz} for comparison. The offset in velocity may be due to the lack of hot halo, which has been shown to reduce the line-of-sight velocities, especially in the LA \citep{Diaz:2011ht}.

\begin{table}[htbp]
	\label{tab:orbit_mw}
   \centering 
   \caption{POSITIONS AND VELOCITIES OF THE CLOUDS}
      \begin{threeparttable}
   \begin{tabular}{@{} lcccccccr @{}} 
  \hline
   Positions & \multicolumn{4}{c}{LMC} & \multicolumn{4}{c}{SMC} \\
   \hline
   Name & x & y & z & $\Delta$R & x & y & z & $\Delta$R \\
   & \multicolumn{4}{c}{(kpc)} & \multicolumn{4}{c}{(kpc)} \\
Observed & $-$1 & $-$41 & $-$28 &-- &15 & $-$38 & $-$44 &-- \\
Besla+12 & $-$1 & $-$42 & $-$26 & 2.2 &6 & $-$39 & $-$35 &12.8 \\
9-1 & 7 & $-$41 & $-$30 & 8.5 &6 & $-$38 & $-$41 & 9.7 \\
7-1 & 10 & $-$36 & $-$35 &13.9 &13 & $-$34 & $-$40 & 6.2 \\
5-1 & 7 & $-$38 & $-$35 &10.6 &7 & $-$38 & $-$34 &12.7 \\
3-1 & 2 & $-$34 & $-$38 &12.1 &2 & $-$35 & $-$38 &15.1 \\
\hline
  Velocities & \multicolumn{4}{c}{LMC} & \multicolumn{4}{c}{SMC} \\
   \hline
Name & V$_x$ & V$_y$ & V$_z$ & $\Delta$V & V$_x$ & V$_y$ & V$_z$  & $\Delta$V \\
   & \multicolumn{4}{c}{(\kms)} & \multicolumn{4}{c}{(\kms)} \\
Observed & $-$79 & $-$227 & 208 & -- &19 & $-$153 & 153 & -- \\ 
Besla+12 & $-$82 & $-$263 & 249 & 54.5 & $-$66 & $-$258 & 198 &142.4 \\
9-1 & $-$1 & $-$347 & 153 & 153.3 & 41 & $-$380 & 156 &227.9 \\
7-1 & 16 & $-$376 & 64 & 227.7 &7 & $-$435 & 116 &285.1 \\
5-1 & 11 & $-$401 & 52 & 249.7 &12 & $-$399 & 55 &264.7 \\
3-1 & $-$93 & $-$312 & 258 & 99.2 &-90 & $-$303 & 259 &213.4 \\

      \hline
      
   \end{tabular}
           \begin{tablenotes}
        	  Note: The origin of the coordinate system is the center of the MW.
        \end{tablenotes}
     \end{threeparttable}
\end{table}

\subsection{Effect of Mass Ratio on the Stream}
\label{subsec:massratio}

In an attempt to understand how the mass ratio between the Magellanic Clouds affects the formation of the Stream, we carried out additional simulations with a 5:1 mass ratio, a 3:1 mass ratio, and a 7:1 mass ratio. This presents an alternative mechanism to strip additional gas from the LMC. In these simulations, we keep the gas disk scale length and mass more similar to the values found in \citetalias{Besla:2012jc}.

These simulations are broadly similar to the features seen in the 9:1 mass ratio case, but differ in the size, shape, and length of the LMC filament. We present the results in \autoref{fig:other_mass_stream} where the three panels show the simulated and observed gas column densities in our three mass ratio simulations.  Although these simulations stripped more mass from the LMC than a 9:1 model with comparable disk scale length, many strip less gas than our fiducial case.

 We also note that in our fiducial model, the Clouds end their orbit very close together on the sky (6 kpc physical separation). This is a trend that continues with all our lower mass ratio simulations. The 3:1 mass ratio interaction merged before the galaxies fell into the MW, and the 5:1 case merged shortly after infall. This is perhaps not surprising given that a 3:1 mass ratio is a major merger and that \citet{Bekki:2009hj} has already hypothesized that a higher mass SMC (8.1\E{10} \msun\ in that case) would not survive in a long-term binary orbit with the LMC.

At present day, our 3:1 mass ratio interaction makes a long thin filament created from the LMC, which lines up on the sky with the SMC filament. The material for this filament is stripped before the Clouds merged, and then stretched across the sky as they fell to their present location. The reason for the alignment between the LMC and SMC filaments is simply because they are falling as one merged galaxy, and given the fact that the real LMC and SMC are currently receding from each other, we do not consider this a plausible formation scenario for the observed filament.

\subsection{Mass Stripping}
\label{subsec:stripping}

\begin{table}[htbp]
\centering
\begin{threeparttable}
\caption{GAS MASS STRIPPING}
\label{tab:mass_stripping}
\begin{tabular}{llllll}
\multicolumn{2}{l}{Name} & M$_{\mathrm{tot}}$\tnote{a}  & M$_{\mathrm{Stream}}$\tnote{b} & M$_{\mathrm{Galaxy}}$\tnote{c} & $f_{\mathrm{stripped}}$\tnote{d} \\
& &  $\times 10^{8}$ \msun & $\times 10^{8}$ \msun &  $\times 10^{8}$ \msun & \\
\hline
& LMC & 5.4 & 0.2 & 5.1 & 4.2\% \\
3:1 & SMC & 9.9 & 3.0 & 6.9 & 29.9\% \\
& Total & 15.3 & 3.2 & 12.1 & 20.9\% \\
\hline 
& LMC & 5.7 & 0.3 & 5.4 & 5.7\% \\
5:1 & SMC & 4.9 & 0.7 & 4.1 & 15.1\% \\
& Both & 10.6 & 1.1 & 9.5 & 10.0\% \\
\hline 
& LMC & 6.7 & 0.2 & 6.5 & 2.7\% \\
7:1 & SMC & 3.5 & 1.6 & 1.9 & 46.8\% \\
& Both & 10.1 & 1.8 & 8.3 & 17.8\% \\
\hline 
& LMC & 12.6 & 2.3 & 10.3 & 18.0\% \\
9:1 & SMC & 5.4 & 1.6 & 3.8 & 29.8\% \\
& Both & 17.9 & 3.9 & 14.0 & 21.6\% \\
\hline 
 & LMC & - & - & 4.4\tnote{e} & -- \\
Observed & SMC & - & - & 4.0\tnote{e} & -- \\
 & Both & 28.4 & 20.0\tnote{f} & 8.4 & --- \\
\hline 

 \end{tabular}
    \begin{tablenotes}
 
          \item [a] Total gas mass at the present day.
	 \item [b] Gas mass stripped from the galaxy (in the Leading and Trailing Streams) at present day.
	 \item [c] Gas mass bound to the galaxy at present day.
	 \item [d] Fraction of gas stripped. 
	   
	\item[e] \HI\ only \citep{Bruns:2005}.
	\item[f] \HI\ and \HII\ \citep{Fox:2014ih}. 
    \end{tablenotes}
\end{threeparttable}
\end{table}

A critical test of any model of the MS is its ability to reproduce the mass contributions to the Stream, Bridge, and LA, while still matching the observed gas mass in the Large and Small Clouds. One of our prime motivations for this work is to explore scenarios where some of this material is stripped from the LMC through an encounter with the SMC. Our 9:1 mass ratio collision strips 3.9\E{8}\msun\ of gas mass - about 20\% lower than the \HI\ observations and an order of magnitude lower than the total gas mass of the Stream. In \autoref{tab:mass_stripping} we present the mass stripped from each Cloud.

The mass budget of the Magellanic System is far more heavily weighted toward the Stream, Bridge, and LA. The total \HI\ mass remaining in the Clouds is 8.4\E{8}\msun\ \citep{Bruns:2005}, with an uncertain amount of ionized gas. The ionized mass is likely on the same order of magnitude given ionization fractions in the Bridge \citep{Barger:2013cv}, but might be less given that most of the H-alpha emission comes from supershells, not diffuse gas \citep{Kennicutt:1995de}. The mass in the Stream, Bridge, and LA is 2$\times$10$^9$ \msun\ in \HI\ and \HII. Further confounding this is that the \HI\ mass of the Stream is likely underestimated given the assumed distance in \citet{Bruns:2005} is closer to the real distance of the LMC and SMC whereas models (including ours) find that a large fraction of the gas is at larger distance.

Even in a high-efficiency stripping scenario, the total gas mass removed from the Clouds is less than the \HII\ mass budget of the Magellanic System and leaves more gas in the Clouds than is observed. Relieving the tension between observed and simulated Stream mass budget, then, is not a matter of adding more gas to the Magellanic Clouds, but to increase the stripping efficiency. Increasing only the gas mass of the Clouds could match the observed mass in the Stream, but at the cost of even greater disparity between the observed and simulated gas masses in the Clouds themselves.

\begin{figure}[htbp] 
   \centering
   \includegraphics[width=3.5in]{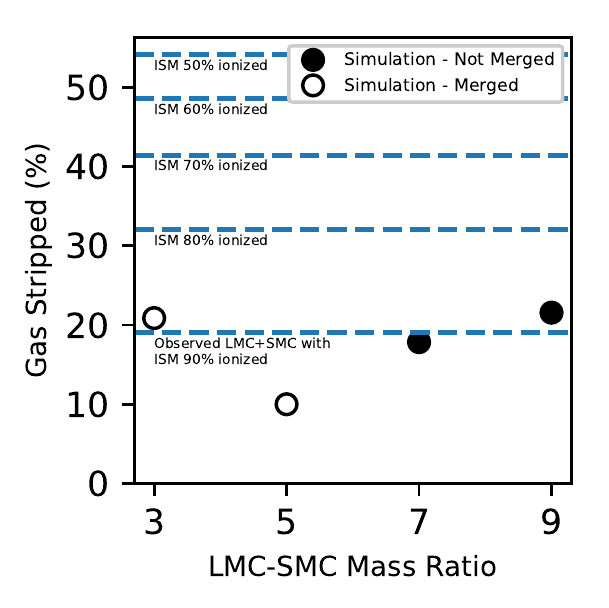} 
   \caption{Percentage of material stripped from both Clouds during their mutual interaction. The dashed lines show the stripping fraction observed in the Magellanic system, computed by taking the ratio of \HI\ and \HII\ in the Stream versus the total \HI\ and \HII\ mass in the two Clouds. Since the \HII\ mass in the clouds is still largely unconstrained, we take several ionization fractions (list under their corresponding line). We also assume that all the material in the Stream was originally in the Clouds. The circles show the stripping percentage in our simulations for different LMC-SMC mass ratios (given by the x-axis). Closed circles are simulations where the two clouds remain separate at the present day, and open circles are simulations where the clouds have already merged by the present day. The clouds would need to have an ionization fraction greater than 80\% to match the stripping in our most aggressive simulation.}
   \label{fig:tot_stripping}
\end{figure}

To illustrate this point we show the total fraction of material stripped from the Magellanic Clouds in \autoref{fig:tot_stripping}. We compare the fraction of gas mass removed from both Clouds during interaction in the various mass ratio simulations to the observed percentage of gas stripped. Different simulations are shown as solid or filled circles, where the solid circles indicate models where the Clouds have not merged by the present day (as is the case observationally). The observed stripped gas percentage is taken by assuming that all the material in the Stream was originally in the Magellanic Clouds and by dividing that mass by the sum of masses in the Stream and both Clouds. For the ionized gas case, we assume a range of ionization fractions in the two Clouds from 50\% (the fraction in the Bridge) to 90\% (the fraction in the outer reaches of the Stream). This range of ionization fractions gives a wide range in stripping fractions shown in several blue dashed lines in \autoref{fig:tot_stripping}. Each line corresponds to a different assumed ionization fraction in the present day Clouds.

Our models range in tidal stripping efficiency from $\sim$ 10\% for the 5:1 mass ratio model to 20\% for the 9:1 model. The mass ratio between the SMC and LMC is uncorrelated with this quantity, because we are showing the combined gas stripping from both galaxies. The changes in disk size, orbit, and mass found in our different simulations all account for some of the difference in removed material. A full search of the orbital configurations where the stripping process is more efficient is beyond the scope of this work (but see the discussion in \citealt{DOnghia:2009kf, DOnghia:2010fk}). This tension necessitates either models where more than 80\% of the gas is removed from the Clouds, or a better understanding of the HII content of the Stream (which could be overestimated). 

Finally, we note again that the Leading Arm observations show a significant lack of gas when compared to the Trailing Stream. Our models show similar masses in both the LA and Stream in all models.

There are good reasons to believe that some material of the Stream was produced from local outflows and by ram pressure stripping of the Clouds gas as they fall into the Milky Way. We discuss these alternative scenarios in \autoref{sec:discussion}. In addition, the \HII\ mass of the Stream (as derived in \citealt{Fox:2014ih}) could include a contribution from cooled coronal gas from the MW halo, not just gas that came from the MCs. Hydro models \citep{Fraternali:2008id} show that cold \HI\ clouds can seed the cooling of the hot halo, thus increasing in mass as time passes. This would reduce the tension in the mass budget between our simulations and the observations.

\section{Discussion}
\label{sec:discussion}

In our model the Magellanic Stream, the most prominent feature of the Southern radio sky and closest example of cold gas accretion, originates from gas removed from the LMC and SMC during their mutual interactions before falling into the MW. The dwarf-dwarf galaxy interaction scenario considered here has been successful at explaining the broad features of the Leading and Trailing Stream components, but predicts the Stream to originate from the SMC and with a mass of order $\sim10^8$ \msun \citepalias{Besla:2012jc}. This is in contrast to the observations showing LMC material in the Trailing Stream and a total gas mass of 2\E{9} \msun \citep{Nidever:2008cz, Fox:2014ih} .

Our models vary the initial size and mass of the Clouds - making the \HI\ disks extended and the galaxies more gas rich in the past. Initial encounters between the Clouds strip material primarily from the SMC, while later interactions also strip material from the LMC. This results in two distinct filaments of gas in the Trailing Stream, mirrored by two components in the Leading Arm. This Trailing Stream material is in qualitative agreement with the metallicity measurements inferred from UV spectroscopy \citep{Fox:2013kc, Richter:2013hk}. In those measurements, there is a filament with 0.1 solar abundances, consistent with an origin in the SMC ~2-2.5 Gyr ago, and a second filament with 0.5 solar abundances, consistent with a recent origin in the LMC. Nevertheless, we have not been able to match the large ionized mass or the high \HI\ column densities in the Trailing Stream. This difficulty is shared by all models of the Stream formation and suggests alternative scenarios, or additional physics may be required to fully align the models and observations.

We stress that the LA region is the critical test for a scenario that envisions the Stream originated by the encounters between the Clouds before they fell into the MW. Thus the LA is predicted to contain material stripped from both the LMC and SMC. Observations that shows multiple metallicities in this region will be the \emph{smoking gun} for the dwarf-dwarf galaxy formation scenario studied here. If, however, the metallicity components in the LA regions point to only a single galaxy origin, the dwarf-dwarf galaxy interaction scenario will require significant updates to match the current observations. Stellar abundances of stars formed in-situ in the LA has already provided evidence of LMC material \citep{Zhang:2017bca}, supporting this scenario, although an alternative scenario (runaway stars) has been presented by \citet{Boubert:2017bc}.

The simulations presented in this work focus on exploring the conditions that lead the LMC to contribute to the origin of the Stream by mutual encounters with the SMC. There are several important caveat to our work, the most important of which is that our simulations do not include a hot gaseous halo for the MW. A hot circumgalactic medium will remove gas from the Clouds due to ram pressure stripping and change the ionization of the Stream due to instabilities \citep{TepperGarcia:2015ho}. Neither of these effects is account for in our models, and both will have the strongest influence on the morphology of the LA \citep{Diaz:2011ht}. 

Even after increasing the total gas content of the Clouds, we cannot reproduce the large supply of gas inferred from absorption measurements. Increasing the gas mass of the Clouds further, without other changes to the models, will not provide a better fit to the data. This is because our models already contain more gas in the Magellanic System than is observed. In our final models, the Clouds are already a factor of $\sim$2 too massive in gas compared to their observed counterparts. To be successful, models need to efficiently remove the gas from both Clouds by mutual gravitational interaction while keeping the galaxies at a separated and intact today.

Given the difficulties reproducing the observed filaments and mass budget of the Magellanic System using only a dwarf-dwarf tidal interaction, we speculate that additional mechanisms might be at play. Since the LMC filament has been traced back to the 30 Doradus star forming region by \citet{Nidever:2008cz} there are good reasons to believe that star formation feedback could play a role in ejecting gas from the LMC. Ram pressure winds might have then swept this hotter, puffier material into a long filament (e.g. \citealt{Mastropietro:2005kv, Diaz:2011ht}). Recent work on this by \citet{Bustard:2018vq} shows how the Milky Way's halo can turn local outflows that would otherwise be fountains into outflows. These outflows, however, are likely not massive enough to explain the missing mass in the Stream.

There has been recent work by \citet{2015ApJ...813..110H} revisiting a purely ram-pressure origin for the trailing arm of the Stream. That work proposed that other dwarfs may have fallen in ahead of the LMC and SMC and created the LA. Although speculative, \citet{Jethwa:2016vv} suggested that the source of these dwarfs could plausibly be the Magellanic Group \citep{DOnghia:2008bu}. Other work investigating the interaction between the LMC and the MW's hot halo by \citet{Salem:2015jn} has found that ram pressure stripping likely only accounts for a few percent of the Stream mass budget. A dense hot halo would cause gas to dissolve rapidly \citep{TepperGarcia:2015ho}, and if the Milky Way's hot halo was denser than n$_h = 10^{-5}$ cm$^{-3}$ at 50 kpc than the survival time would be less than 500 Myr \citep{Murali:2000kz}. In this case, the SMC filament in our model (which is older than 2 Gyr) would be completely evaporated. Finally, ram pressure could explain why there seems to be a disparity found between the density of gas in the Leading Arm, which would be exposed to a strong ram pressure effect, and the gas found in the Trailing Stream.

If the clouds entered as a group (see e.g. \citealt{DOnghia:2008bu}) then they may have brought a significant amounts of hot gas in their circumgalactic medium \citep{Bordoloi:2014jo}. This gas would also interact with the Milky Way's hot halo and be preferentially swept to the Trailing Stream (Pardy et al. in prep)

Instead of one cause, a combination of tidal interaction, outflows, and ram pressure is probably required to fully match the Stream location and mass. Testing these theories is complicated by the uncertain chemical enrichment histories of the Clouds and how strong their metallicity gradients are (for instance, see \citealt{ToribioSanCipriano:2017ix} for evidence that the metallicity gradients in both galaxies are flat). Any gas filaments produced by outflows would come from the inner material, and may be more enriched, while gas stripped during an interaction would come from outer material. Further confounding this is that, in our model, the gas was stripped anywhere from one to a few Gyrs ago, making it likely to have lower metallicities than the present day LMC or SMC material. 

Additional passes between the MW and the MCs would make the ram pressure origin more plausible and even aid in the tidal stripping due to the MW. Indeed, updated models of the MW's rotation curve \citep{2009MNRAS.392L..21S, Diaz:2011bb}, accurate treatment of the MW response to the LMC \citep{2015ApJ...802..128G}, and three-year proper motion measurements that have decreased since the first data release \citep{Kallivayalil:2013vq} have since marginally increased the likelihood of a second passage scenario. We note, however, that any past pericenters were likely further from the Clouds' present positions, and they are likely the closest they have ever been to the MW, making any of these effects relatively minor.

Future observations will be critical to further refine our theory of the Magellanic System formation. On going work by the WHAM team on observations of the MS and Clouds in \halpha\ should provide the total gas mass of the System, especially the gas that has been stripped from the Clouds (see \autoref{subsec:stripping}). 

Recent ultraviolet studies have shown the chemical abundances in the LA are spatially variable, with O/H ranging from 4--6\% solar to 29\% solar (Fox et al. 2018; Richter et al. 2018, see also \citealt{Lu:1998gr}) in the clumps known as LA II and LA III. This range covers both SMC-like and LMC-like abundance patterns (given that both galaxies had lower metallicities in the past than they do now), suggesting that both galaxies contributed to the generation of the LA, as predicted by our tidal simulations. Continued absorption measurements of the Stream and LA are needed to reveal the full enrichment history of the Magellanic System.

\section{Conclusions}
\label{sec:conclusions}

We have presented a model of multiple encounters between the LMC and the SMC that can produce bifurcations in the Trailing and Leading Arms of the MS. Our models assumed a more gas rich SMC in the past and an LMC with a more extended gaseous disk. These assumptions increase the efficiency and the amount of gas tripped by mutual gravitational interactions between the LMC and the SMC before they fell into the MW potential. Our results can be summarized as follows: 
\begin{itemize}
\item Increasing the gas content in the SMC and LMC and the significantly increasing their scale lengths  can strip additional material from the disk of the LMC and create two filaments of \HI\ gas in the Trailing Stream - in qualitative agreement with observations. 
\item Increasing the SMC mass increases the probability that the Clouds would have merged before the present day. However, even with increase of gas stripped from the two Clouds, we are unable to reproduce the mass budget of the observed Magellanic System, where the majority of mass is presently located in neutral or ionized gas in the Stream, and very little remains in the Clouds themselves.
\item The LA shows material from both the LMC and the SMC. If the Stream was formed in a dwarf-dwarf interaction scenario similar to the one laid out in this paper, then future observations of the LA should uncover evidence of metallicity from both Clouds. Recent ultraviolet absorption measurements of spatial variation in chemical abundances in the Leading Arm, which show that gas from multiple sources is present, although the nature is still uncertain \citep{Fox:2018vz}.
\item Increasing the gas mass of the LMC and SMC progenitors in our models has not solved the mass budget problem or the mismatch between models and observations in the Stream column density. This suggests additional physics or sources of gas are important in creating the Stream.
\end{itemize}

\acknowledgments

SP thanks David Nidever for providing data from his 2010 work. ED gratefully acknowledges the hospitality of the CCA at the Flatiron Institute during the completion of this work. This research made extensive use of many open-source Python packages including Astropy, a community-developed core Python package for Astronomy (Astropy Collaboration, 2013), the Pathos\footnote{http://trac.mystic.cacr.caltech.edu/project/pathos} multiprocessing library \citep{McKerns:2012wd}, and the Scipy library \citep{Perez:hy, Hunter:ih, vanderWalt:dp}. 

\bibliography{bibtex_library}
\clearpage

\end{document}